\documentclass[]{spie}  %>>> use for US letter paper
\usepackage{amsmath,amsfonts,amssymb}
\usepackage{graphicx}
\usepackage[colorlinks=true, allcolors=blue]{hyperref}
\usepackage{subcaption}
\usepackage{wrapfig} % dans le préambule
\usepackage{float}

\title{Optical concept model of the future cosmology project BISOU}

\author[a,b]{*Morgane Loquet Le Gall}
\author[a]{Bruno Maffei}
\author[a]{Pierre Guiot}
\author[c]{Creidhe O'Sullivan}
\author[c]{Neil Trappe}
\affil[a]{Institut d'Astrophysique Spatiale, Université Paris-Saclay, CNRS, Building 121, Orsay, France}
\affil[b]{Centre national d'études spatiales (CNES), 75039 Paris, France}
\affil[c]{Department of Physics, Maynooth University, Co. Kildare, Ireland}

\authorinfo{Further author information: (Send correspondence to M. Loquet Le Gall) \\ E-mail: morgane.loquet-le-gall@universite-paris-saclay.fr}

\begin{document} 
\maketitle

\begin{abstract}
We present an optical analysis of BISOU (Balloon Interferometer for Spectral Observations of the primordial Universe), an astronomical balloon-borne pathfinder spectrometer developed as part of a preparatory study for a future space mission aiming at measuring spectral distortions of the cosmic microwave background (CMB). The BISOU optical system is based on a differential polarising Fourier Transform Spectrometer (FTS) that receives inputs from both a sky-facing telescope and an internal calibration source. The FTS focal planes are equipped with bolometric detectors coupled to multimode feed horns, with distinct focal planes dedicated to the low (90~-~300~GHz) and high (0.3~-~1.5~THz) frequency bands. The optical analysis first relies on ray-tracing simulations to establish the overall configuration of the system, before proceeding to more advanced Gaussian beam and physical optics analyses.
\end{abstract}


\section{INTRODUCTION}
\label{sec:intro}  % \label{} allows reference to this section

The spectrum of the Cosmic Microwave Background (CMB) was measured in 1991 by the COBE-FIRAS satellite, revealing it to be isotropic and very close to a perfect blackbody emission at a temperature of 2.725 K. \cite{FIRAS} However, theory predicts that small deviations from a pure Planck's law spectrum, known as spectral distortions, are expected. These distortions carry valuable information about the thermal history of the Universe and the physical processes that occurred during its earliest epochs, and their measurement is one of the major challenges in cosmology today \cite{spectral_distortions}. FIRAS limited sensitivity only set an upper limit, driving the community to propose new space missions with a much higher sensitivity, such as PIXIE \cite{PIXIE} and FOSSIL \cite{FOSSIL}. BISOU (Balloon Interferometer for Spectral Observations of the Universe), a balloon-borne spectrometer proposed as a pathfinder for a future dedicated space mission, is now in a Phase A CNES study. BISOU \cite{BISOU} will be able to measure the y-monopole of spectral distortions for the first time, and will improve on the knowledge of the Cosmic Infrared Background (CIB) \cite{CIB}. Understanding this instrument and characterising any systematic effects is critical to reaching the required sensitivity for such measurements. In particular, we aim to minimize the efficiency losses along the optical path, notably by identifying their origin. In this paper, we focus specifically on the impact of the Fourier transform spectrometer moving mirror, forming the basis of our instrument.

\newpage

\section{The BISOU instrument concept}

\label{sec:instrument concept}

\begin{figure}[!h]
    \centering
    \includegraphics[width=1\linewidth]{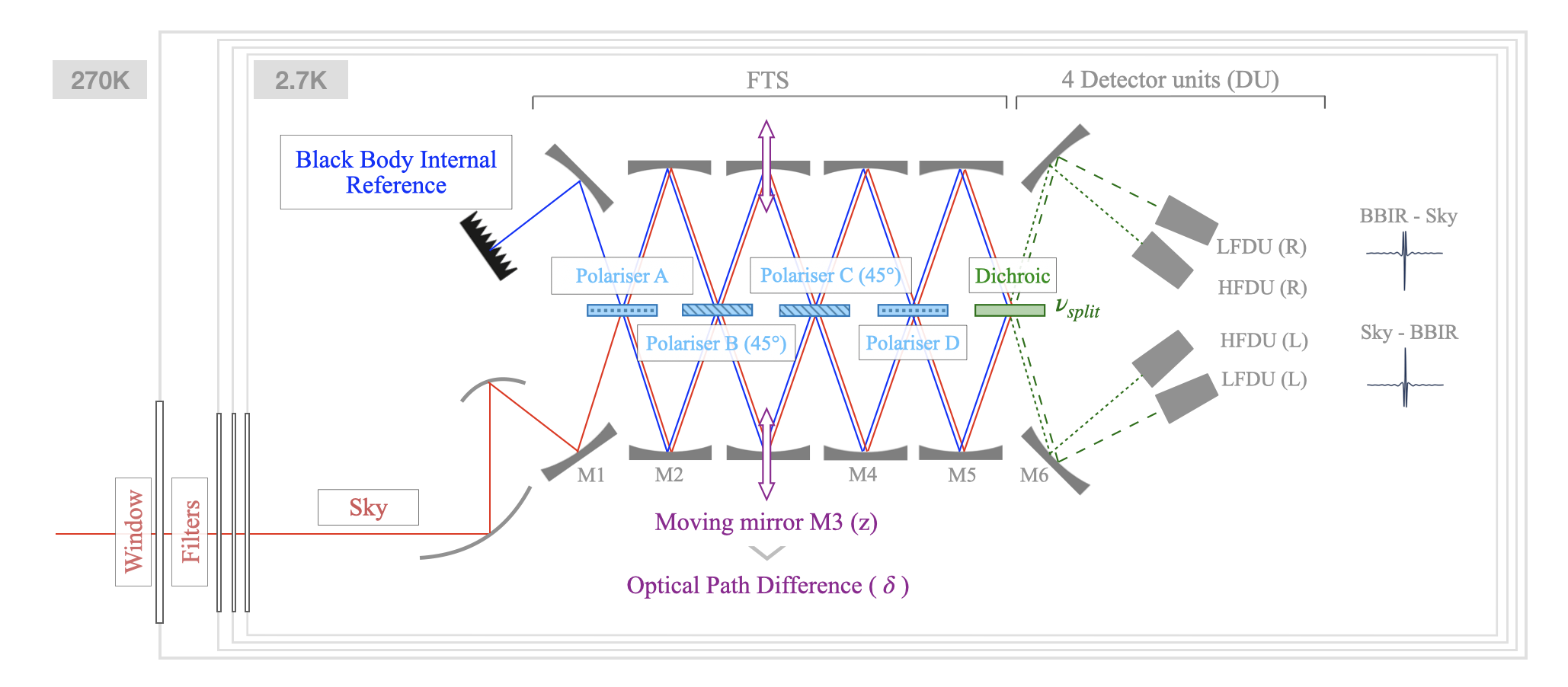}
    \caption{Schematic diagram of the BISOU instrument. One beam comes from the sky and another from the internal 2.7~K reference. An optical path difference is introduced by the pair of moving mirrors. Polarisation is split using polarising filters, and then recombined to produce the desired interference pattern. Before the focal plane, a dichroic separates the high and low frequencies, producing four focal planes. Each detector will be coupled to a multimode feedhorn optimized for its respective frequency band.}
    \label{fig:concept bisou}
\end{figure}

The design of the BISOU instrument is inspired by the PIXIE proposal \cite{PIXIE}, but it has been adapted to meet the constraints of a balloon-borne experiment rather than those of a space mission. The instrument is a differential Fourier Transform Spectrometer (FTS) based on a polarised Martin–Puplett interferometer \cite{MPInterferometer} with two inputs: one arm observes the sky, while the other is looking at an internal blackbody reference maintained at the temperature of the CMB (2.7~K). The FTS output, where detectors are located, collects a signal consisting of a constant part and a FTS scan-modulated part proportional to the difference between the two inputs. The beams are split and recombined using wire-grid polarisers, and an optical path difference is introduced by a pair of moving mirrors. The outputs of the FTS are coupled to detectors with multimode feedhorns. The accurate reference is essential for spectral distortion measurements, as the signal corresponds to tiny deviations from an almost perfect blackbody spectrum. Without a well-characterized calibrator, systematic effect will limit the ultimate sensitivity. The overall concept of the instrument is illustrated in Figure \ref{fig:concept bisou}

As the instrument is designed for a balloon mission, it faces additional challenges compared to a traditional space mission. This is mainly due to the residual atmosphere that is still present at an altitude of 40~km (P~=~3~mbar). In order to achieve the required sensitivity, the instrument will be maintained at 2.7~K, while the focal plane will be cooled to a few hundred millikelvin in order to reach the operating temperature of the detectors. Therefore, the presence of the atmosphere necessitates the instrument being enclosed in a cryostat capable of withstanding the pressure difference between the external environment and the internal vacuum.

The photon noise is dominated by the thermal emission of the residual atmosphere and the warm optical component (270~K) such as the window and filters. While the window is transparent to millimiter and sub-millimeter waves and is here to close the cryostat, the filters limit the thermal load on each temperature stage of the instrument. All these optical components must be carefully characterised in order to separate and subtract their contributions from the measured signal. \cite{thesexavier} 

To improve the sensitivity of the instrument, the wide frequency band is subdivided using a dichroic filter. Frequencies below the cut-off frequency (300~GHz \cite{thesexavier}) are transmitted through the dichroic, while higher frequencies are reflected by its surface. Separating frequencies into two sub-bands significantly reduces the power received by the detectors and, consequently, the photon noise in the low-frequency band. This is particularly advantageous for a balloon-borne mission, where high-frequency emissions are largely caused by the internal emission of warm instrument components. This reduction in photon noise improves the sensitivity and overall performance of the instrument, particularly in the low-frequency channel. \cite{spiexavier}

The instrument's key mechanism relies on a pair of moving mirrors, the displacement of which is directly linked to the instrument’s specifications. The maximum optical path difference (OPD) between the two arms determines the instrument's spectral resolution, which is set to $\Delta \nu~=~15~\text{GHz}$ according to equation \ref{eq:OPDmax}. Furthermore, the optical path difference spacing  $\delta$ between two measurement points in the interferogram relates to the maximum frequency of the BISOU broadband signal, as described by equation \ref{eq:dOPD}.

\begin{equation}
\label{eq:OPDmax}
    OPD_{max} = \frac{c}{\Delta \nu}
\end{equation}

\begin{equation}
\label{eq:dOPD}
    \delta = \frac{c}{2~\nu_{max}}
\end{equation}

As the mirrors are displaced from their nominal positions by a distance $z$ through the FTS mechanism (figure~\ref{fig:concept bisou}), an additional optical path length of approximately $2z$ is introduced in one arm, while the path length in the other arm is reduced by the same amount. This results in an optical path difference of approximately $\delta~=~4z $. Further investigation using ray-tracing simulations is required to establish a more accurate and comprehensive relationship between mirror displacement and the actual optical path difference. With BISOU’s spectral resolution of $\Delta \nu~=~15~\text{GHz}$, we require a maximum optical path difference of 2~cm, which corresponds to a total mirror stroke of $\pm ~2.5~\text{mm}$. In this paper, we explored the case of a mirror stroke of $\pm~10~\text{mm}$, corresponding to four times the stroke required to reach our target sensitivity.

A cryogenic Breadboard model (BBm) is being built at the Institut d'Astrophysique Spatiale (IAS). This cryogenic prototype represents an important step toward validating the concept, studying systematic effects that are still unknown for this type of instrument, and testing new technologies such as detectors. In particular, the asymmetry of the two optical paths will need to be studied.

\section{Optical Modelling}

The initial optical design of the instrument was carried out using ray-tracing in the industry-standard Zemax OpticStudio \cite{zemax} and GRASP \cite{grasp} software. The more detailed analyses, especially at the lower operating frequencies where the beam is the largest ($\nu_{min}~=~90~\text{GHz}$ or $\lambda_{max}~=~3.3~\text{mm}$), were carried out using full vector physical optics in GRASP. The beam is well-modelled by rays at the high-frequency end of the band.
    
\subsection{Ray Tracing}

\begin{figure}[!h]
    \centering
    \includegraphics[width=0.85\linewidth]{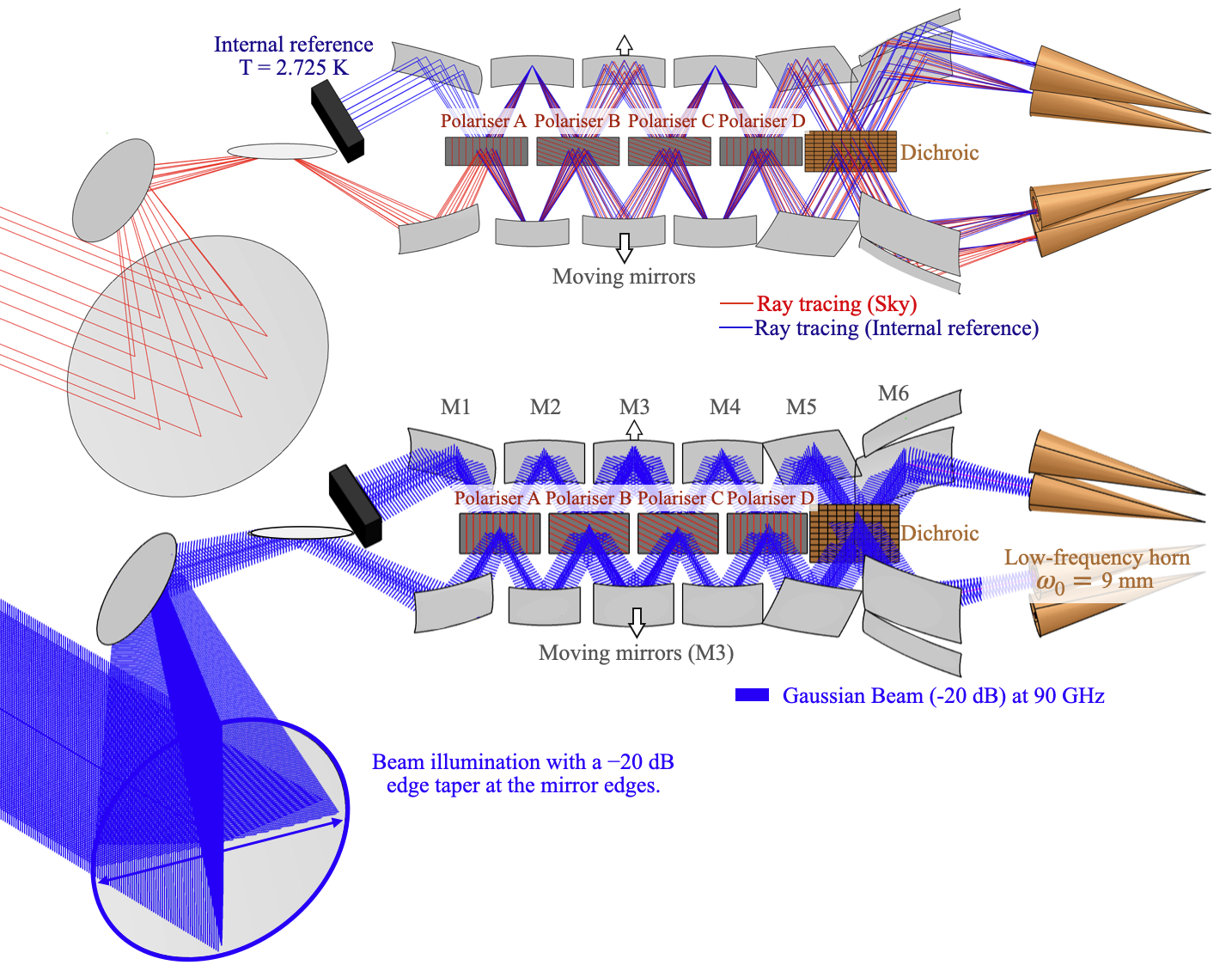}
    \caption{Top : View of the BISOU instrument with ray tracing. Bottom : A 90 GHz Gaussian beam with a 9 mm waist is propagated through the instrument thanks to the Ticra GRASP software \cite{grasp}. The beam envelope is set at -20~dB to show the edge taper of the primary mirror's edge. At low frequencies, the beam is strongly affected by diffraction, whereas at high frequencies it behaves closer to the ray-tracing limit }
    \label{fig:design}
\end{figure}

The initial optical design of the current version of our instrument was based on geometrical optics (GO) (figure~\ref{fig:design}, top panel). This approximation is valid when rays are considered infinitely thin, the light source is point-like, and optical surfaces are ideal (with a reflection coefficient of 1) \cite{opticgeo}. Ray tracing is particularly useful for multi-wavelength instruments, as the behavior of rays is independent of wavelength due to the achromatic nature of our reflective optics.

This approximation allowed us to quickly and easily model light propagation through the instrument, thanks to the Zemax software \cite{zemax}. It helped us select the most suitable mirror types for our needs, to ensure the beam passes through specific optical elements. In our BISOU design, the last FTS mirror M5 has been changed from an elliptical mirror, used in PIXIE design, to a parabolic mirror. This produces a collimated beam on the dichroic, ensuring that all incident rays reach it at the same angle of incidence, which simplifies the characterization of this critical component. The mirror M6 is now used to focus the beam onto the feedhorns. To achieve the detectors’ operating temperature of a few hundred millikelvin, the detectors and their associated feedhorns must be co-located within the same area. 

\begin{figure}[!h]
    \centering
    \includegraphics[width=0.8\linewidth]{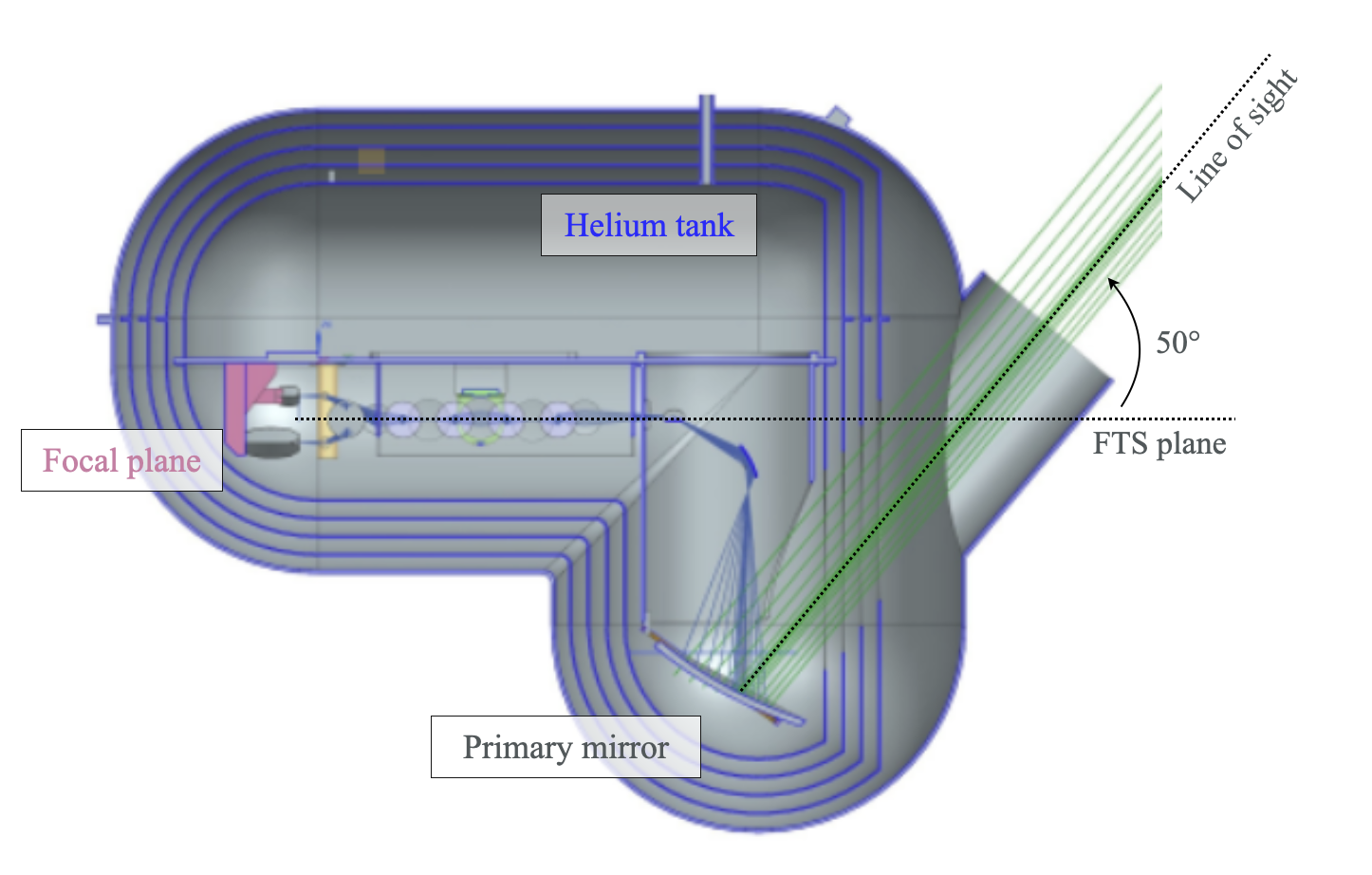}
    \caption{Cross-sectional view of the BISOU CAD model. The helium tank is visible above the cold plate, on which the FTS is mounted. The line of sight is highlighted, at an angle of 50° with respect to the FTS plane.}
    \label{fig:cad}
\end{figure}

As an example, one of the most important constraints for the BISOU mission is related to its balloon-based platform. The instrument needs to fit in a standard gondola (CARMEN~-~CNES \cite{carmen}), the lift resulting from a 800~000~$m^3$ He balloon. The BISOU FTS design (mirrors M1 to M5), has been simplified to a five-mirror configuration compared to the six-mirror design used in instruments such as PIXIE original design. The mirror M1 directly images the internal reference, thereby avoiding the need for a complex and heavy optical path to inject the reference signal into the FTS arm. This reduction in the number of optical elements enables a decrease in the overall mass of the FTS while preserving the same interferometric signal formulation at the detector level. 

We have chosen a telescope following the Mizugushi-Dragone (MD) condition \cite{MD_condition}, in its Gregorian version, in order to minimize the cross-polarisation and the astigmatism of the system. Its output had to be modified to be used as input of the interferometer to illuminate the first of the five concave mirrors. To avoid any risk of field of view vignetting by the balloon we considered a limiting elevation angle of 50° compared to the plane of the FTS horizontal plane. This constraint is added to the MD condition and complexifies the design of the telescope. A view of the instrument within its cryogenic environment is shown in Figure \ref{fig:cad}.

\subsection{Gaussian beam}

In the millimetre-wave domain, Gaussian optics is preferred to the usual GO approach when the size of the system (mirrors, aperture, ...) is comparable to the wavelength (few mm). \cite{goldsmith} In this regime, the effects of diffraction, the beam divergence or the phase curvature become critical and are not modelled by straight rays. Even if BISOU will operate with multimoded horn, the first modelling is performed by considering them as monomode gaussian beams. Gaussian optics allows to model the size of the beam quickly and its behaviour across the instrument, which is critical in order to determine the size of the instrument and begin to optimise the optical system and the overall design.  

In our first modelling, we consider that the beam is a gaussian, entirely described by two parameters, the frequency $\nu$ and the size of the waist $\omega _0$. The waist of a gaussian beam is its minimum beam radius, corresponding to the region where the wavefront is nearly planar and the phase most uniform. It defines the effective size of the equivalent Gaussian source and sets the beam’s divergence during propagation. As an approximate starting point typical at those frequencies, we set our feed's waist at 9~mm at 90~GHz. All further modelling is performed at 90~GHz, the lowest frequency in the BISOU frequency range, where the beam is the widest.

Thanks to this gaussian feed, we have a simple analytical way to compute the size of the beam at each optical element without running a simulation. We aim at a edge taper of -35~dB for each edge of the mirror in the FTS in order to limit spillover and guarantee that the useful field remains compatible with the detector’s field of view. Such a level of edge attenuation ensures that 98\% of the beam power is contained inside the envelope of the beam, strongly reducing diffracted contributions that could degrade the instrument’s response. This choice must, however, be interpreted in light of the horn transmitter/receiver convention: the pattern used to describe the mirror illumination actually corresponds to the horn’s radiation pattern in transmission, which is identical to its reception response by electromagnetic reciprocity. However on the main reflector we aim for a edge taper of -20~dB to lighten the design constraints and keep a reasonably sized primary mirror. The gaussian beam is an effective way to adjust mirror parameters (reflection angles, focal distance...) in order to reach the target illumination on each of the optical elements. This optimization allowed us to reduce the size of mirrors M2 and M4 compared to the other mirrors of the FTS. While the largest FTS mirror have a side length of 90~mm, we were able to decrease the side length of the smaller mirrors to 70~mm without increasing the spillover. The bottom panel in Figure \ref{fig:design} shows the 90~GHz Gaussian beam as it propagates through the optical system.

\section{PHYSICAL OPTICS SIMULATION}

We used GRASP to study how displacements of the moving mirror change the mirrors illumination and impact the instrument’s performance. In particular, we examined the instrument’s spillover at the extreme mirror positions, as well as any beam deformation throughout the optical system.

\subsection{Spillover}

We computed the spillover throughout the optical system using the standard antenna emission/receiving convention. By shifting the moving mirror to its extreme position, we can analyze the power radiated by the feedhorns that falls outside the optical elements. In this paper, we considered a maximum mirror stroke of $\pm$~10~mm, which exceeds  the displacement likely to be used on BISOU. In the reception convention, this lost power corresponds to undesirable radiation that the feedhorn would see from external elements along the optical path. This approach allows us to quantify the efficiency and identify potential sources of stray light at each stage of the system.

This simulations were performed using gaussian beams. This approach was chosen because the gaussian beam is a predefined source model available in Ticra GRASP \cite{grasp}, and is thus easy to configure and implement. It also allows for fast computations at each frequency, which is particularly advantageous for the preliminary analyses. This initial modelling provides a good way to estimate the changes in the beam when the moving mirror is displaced.

\begin{figure}[!h]
    \centering
    \includegraphics[width=0.9\linewidth]{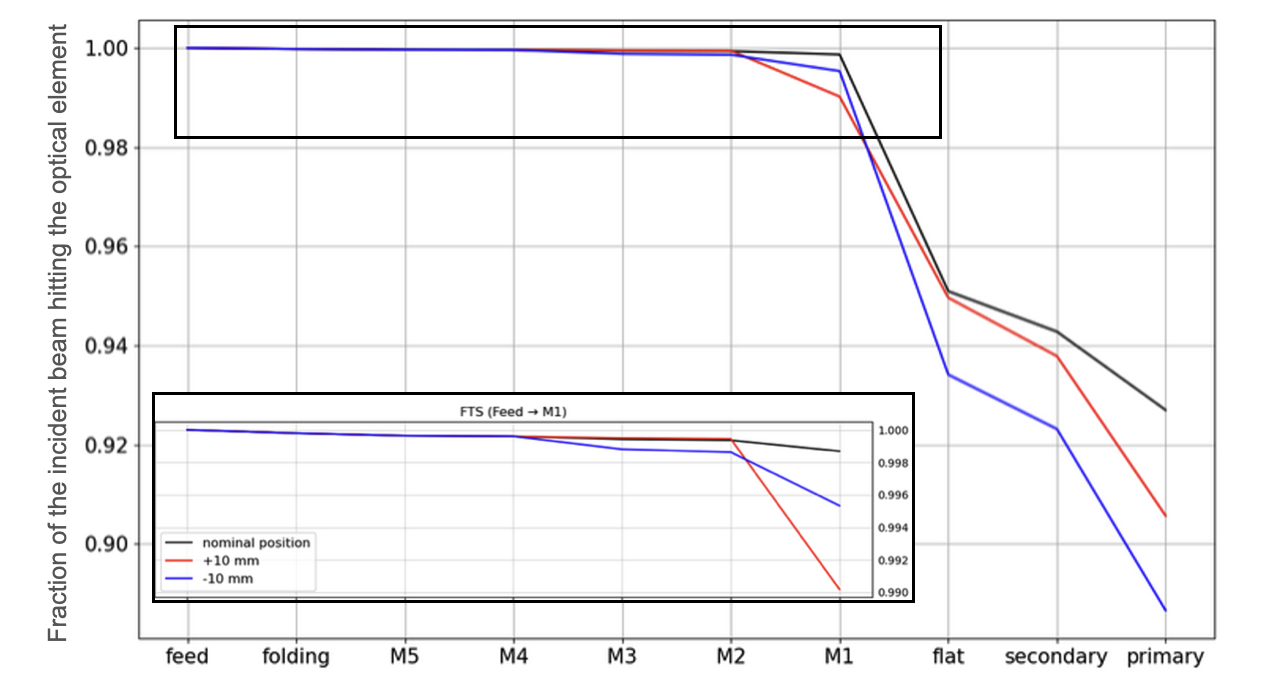}
    \caption{A comparison of the total instrument spillover for three different mirror positions. We chose to compute this spillover using the emission convention (from the focal plane to the sky). We calculated the fraction of the initial beam that reach each optical element. A value of 1 corresponds to the power emitted by the horn at its aperture.}
    \label{fig:spillover}
\end{figure}

Thanks to our preliminary design using the analytic Gaussian beam with the –35~dB envelope, we keep the power loss through the FTS (feed to M1) below 0.2\% at 90~GHz when the moving mirror is in its nominal position. The total power lost through the system (FTS + telescope) remains below 7.5\%. The results are shown in Fig \ref{fig:spillover}.

We also calculated the power lost through the optical system when the moving mirrors M3 were moved to their extreme positions ($\pm$~10~mm). Due to the -20~dB edge taper requirement on the sides of the telescope's primary mirror, displacement of the mirror and the difference between the original and modified optical paths does not dramatically affect the spillover of the instrument and the loss stay around 10\%. We note, however, an asymmetry between the optical paths at +10~mm and -10~mm mirror stroke, likely due to the mirror shapes. This asymmetry will need to be investigated more thoroughly in future simulations. Nevertheless, the results presented here correspond to a resolution four times higher than the one required for BISOU, with a mirror stroke four times larger than needed. These results are therefore very encouraging for future developments.

\subsection{Instrument performance}

When the moving mirrors are displaced, the rays no longer strike the centres of the mirrors, which can introduce optical aberrations. To verify that these aberrations remain acceptable, we use a Gaussian beam to simulate the system and assess how the beam shape changes as the mirror moves. This ensures that the optical performance of the instrument is maintained throughout the full range of mirror displacement.

\begin{figure}[!h]
    \centering
    \includegraphics[width=1\linewidth]{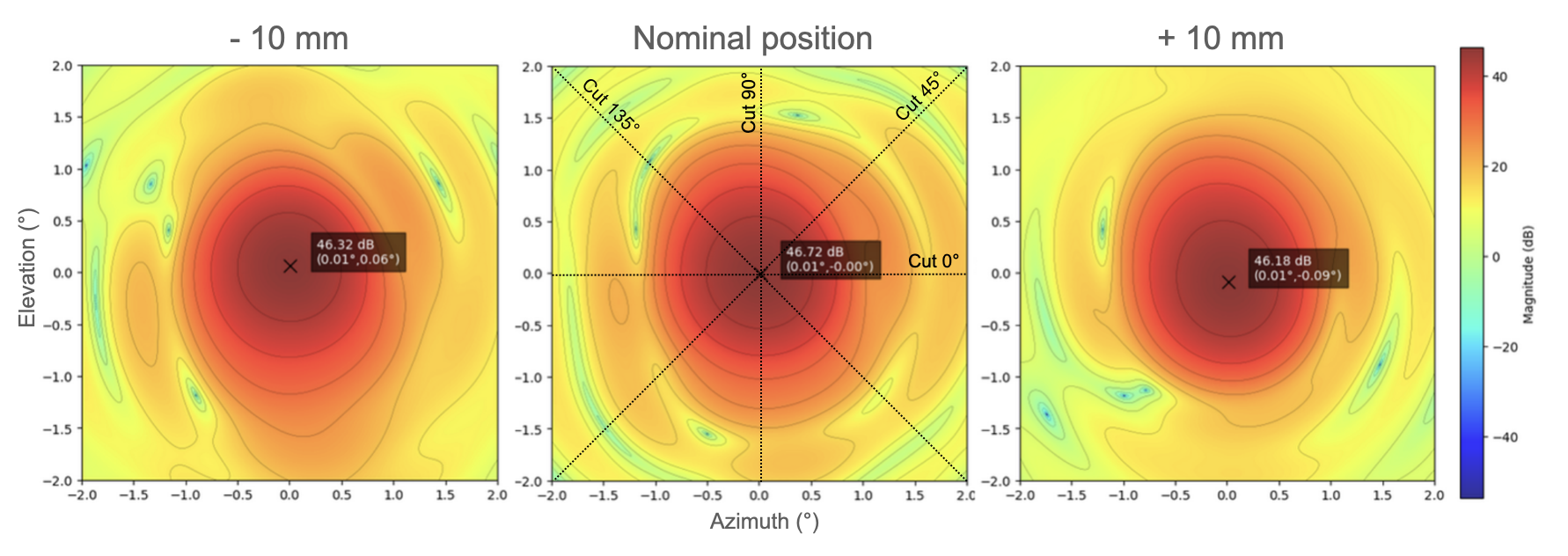}
    \caption{Simulated beam patterns on the sky for three different positions of the moving mirror. The leftmost figure corresponds to the -10 mm position, meaning that both mirrors are shifted 10 mm to the left from their default position. The middle figure corresponds to the default position, and the figure on the right corresponds to the +10 mm position, where both mirrors are shifted to the right. The colour scale represents the beam magnitude in dB. The black cross marks the beam's peak, and the corresponding magnitude and angular coordinates are shown in the labels. The angles of the different cuts used in Figure \ref{fig:antenna_diagram} are illustrated in the central figure. The results demonstrate that the beam remains centred and retains its shape across the different mirror positions, with only slight variations in the peak direction at 90~GHz.}
    \label{fig:beam mirror moving}
\end{figure}

\begin{figure}[!h]
    \centering
    \includegraphics[width=1\linewidth]{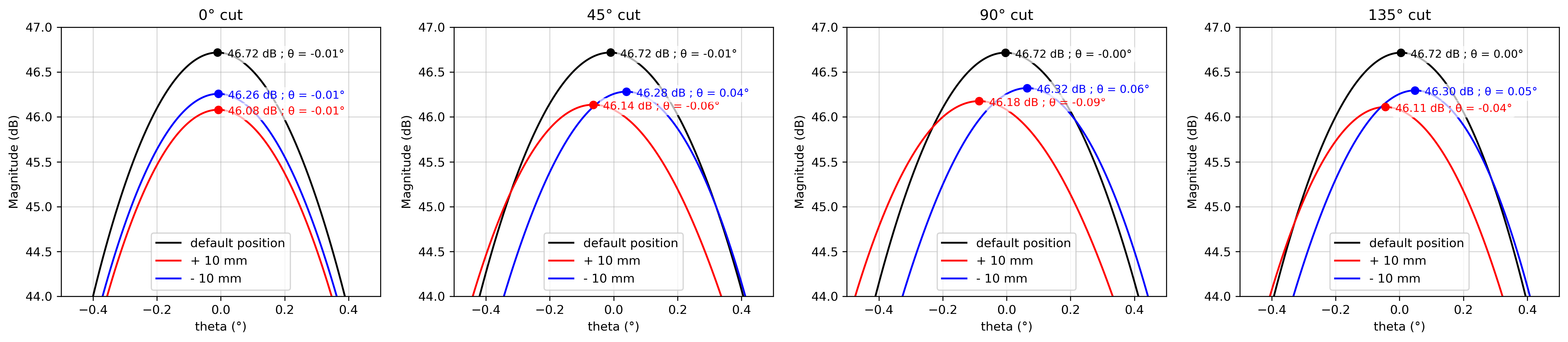}
    \caption{Radiation patterns (magnitude in dB) as a function of the angle $\theta$ for different angular cuts (0°, 45°, 90°, and 135°). Results are shown for three positions of the moving mirror: nominal position (black), $+$10 mm displacement (red), and $-$10 mm displacement (blue). A slight shift in the main lobe direction and variations in the peak magnitude can be observed depending on the positional offset.}
    \label{fig:antenna_diagram}
\end{figure}

 Results of the simulations, shown in figure \ref{fig:beam mirror moving} and \ref{fig:antenna_diagram} show that the beam remains essentially aligned along the same horizontal axis, with minimal variation in the vertical axis. Depointing remain below ± 0.1°. In addition, the overall gaussian shape of the beam is preserved as well as the directivity. The maximum directivity loss between the nominal position and +~10~mm remains limited to 0.5~dB.  This indicates that the pointing stability of the system is very robust to mirror movements within the considered range.

%The effect of the instrument optics is first evaluated using an ideal beam, perfectly Gaussian and initially free of cross-polarisation. This approach allows us to isolate and quantify the cross-polarisation generated progressively during propagation through the instrument, in particular between the polarising elements of the FTS.

\section{CONCLUSION}

In this paper, we have described the different approaches used in the design of the BISOU instrument. The ray tracing and gaussian beam approaches were employed to accommodate the constraints associated with the balloon-based nature of the mission while taking the beam width into account. 
This design enabled us to compute the beam through the optical system and estimate the effect of the movement of the moving mirror on the projected beam. 
In future work, we will model the behaviour of the multimode horn throughout the system and compute a more realistic beam pattern. In addition, a breadboard model of the instrument being constructed at IAS before the end of the year will enable us to perform the first measurements and compare them with our model.

\acknowledgments % equivalent to \section*{ACKNOWLEDGMENTS}       
 
The authors acknowledge financial support from the Centre national d’études spatiales (CNES), France (ROR: https://ror.org/04h1h0y33), within the framework of the BISOU balloon mission.

This research is supported by R\'egion \^Ile-de-France through fundings with reference IDF-DIM-ORIGINES-2023-4-07 and IDF-DIM-ORIGINES-2024-1-06

% References
\nocite{*}
\bibliography{report} % bibliography data in report.bib

@ARTICLE{FIRAS,
       author = {{Mather}, J.~C. and {Fixsen}, D.~J. and {Shafer}, R.~A. and {Mosier}, C. and {Wilkinson}, D.~T.},
        title = "{Calibrator Design for the COBE Far-Infrared Absolute Spectrophotometer (FIRAS)}",
      journal = {The Astrophysical Journal},
         year = 1999,
        month = feb,
       volume = {512},
       number = {2},
        pages = {511-520},
          doi = {10.1086/306805},
archivePrefix = {arXiv},
       eprint = {astro-ph/9810373},
 primaryClass = {astro-ph},
       adsurl = {https://ui.adsabs.harvard.edu/abs/1999ApJ...512..511M}
}

@ARTICLE{pixie,
       author = {{Kogut}, Alan and {Aghanim}, Nabila and {Chluba}, Jens and {Chuss}, David T. and {Delabrouille}, Jacques and {Dvorkin}, Cora and {Fixsen}, Dale and {Ghosh}, Shamik and {Hensley}, Brandon S. and {Hill}, J. Colin and {Maffei}, Bruno and {Pullen}, Anthony R. and {Rotti}, Aditya and {Sabyr}, Alina and {Switzer}, Eric R. and {Thiele}, Leander and {Wollack}, Edward J. and {Zelko}, Ioana},
        title = "{The Primordial Inflation Explorer (PIXIE): mission design and science goals}",
      journal = {Journal of Cosmology and Astroparticle Physics},
         year = 2025,
        month = apr,
       volume = {2025},
       number = {4},
          eid = {020},
        pages = {020},
          doi = {10.1088/1475-7516/2025/04/020},
archivePrefix = {arXiv},
       eprint = {2405.20403},
 primaryClass = {astro-ph.CO},
       adsurl = {https://ui.adsabs.harvard.edu/abs/2025JCAP...04..020K}
}

@INPROCEEDINGS{BISOU,
       author = {{Maffei}, Bruno and {Aghanim}, Nabila and {Aumont}, Jonathan and {Battistelli}, Elia and {Chluba}, Jens and {Coulon}, Xavier and {De Bernardis}, Paolo and {Douspis}, Marian and {Grain}, Julien and {Hill}, J.~C. and {Kogut}, Alan and {Kuruvilla}, Joseph and {Lagache}, Guilaine and {Macias-Perez}, Juan and {Masi}, Silvia and {Matsumura}, Tomotake and {Mele}, Lorenzo and {Monfardini}, Alessandro and {O'Sullivan}, Cr{\'e}idhe and {Pagano}, Luca and {Pisano}, Giampaolo and {Ponthieu}, Nicolas and {Remazeilles}, Mathieu and {Ritacco}, Alessia and {Rotti}, Aditya and {Savini}, Giorgio and {Sauvage}, Valentin and {Shitvov}, Alexey and {Stever}, Samantha L. and {Tartari}, Andrea and {Thiele}, Leander and {Trappe}, Neal and {Aubrun}, Jean-Fran{\c{c}}ois and {Laurens}, Andr{\'e} and {Pheav}, Dominique and {Vacher}, Fran{\c{c}}ois},
        title = "{BISOU: a balloon project for spectral observations of the early universe}",
    booktitle = {Millimeter, Submillimeter, and Far-Infrared Detectors and Instrumentation for Astronomy XI},
         year = 2022,
       editor = {{Zmuidzinas}, Jonas and {Gao}, Jian-Rong},
       series = {Society of Photo-Optical Instrumentation Engineers (SPIE) Conference Series},
       volume = {12190},
        month = aug,
          eid = {121900A},
        pages = {121900A},
          doi = {10.1117/12.2630136},
       adsurl = {https://ui.adsabs.harvard.edu/abs/2022SPIE12190E..0AM}
}

@ARTICLE{MPInterferometer,
       author = {{Martin}, D.~H. and {Puplett}, E.},
        title = "{Polarised interferometric spectrometry for the millimeter and submillimeter spectrum.}",
      journal = {Infrared Physics},
         year = 1970,
        month = jan,
       volume = {10},
        pages = {105-109},
          doi = {10.1016/0020-0891(70)90006-0},
       adsurl = {https://ui.adsabs.harvard.edu/abs/1970InfPh..10..105M}
}

@BOOK{opticgeo,
       author = {{Born}, M. and {Wolf}, E.},
        title = "{Principles of Optics Electromagnetic Theory of Propagation, Interference and Diffraction of Light}",
         year = 1980,
       adsurl = {https://ui.adsabs.harvard.edu/abs/1980poet.book.....B}
}

@misc{zemax,
  author       = {{Ansys}},
  title        = {Ansys Zemax OpticStudio},
  year         = {2024},
  url          = {https://www.ansys.com/products/optics/ansys-zemax-opticstudio},
}

@misc{grasp,
  author       = {{TICRA}},
  title        = {GRASP Software},
  year         = {2024},
  url          = {https://www.ticra.com/software/grasp},
}

@ARTICLE{MD_condition,
       author = {{Dragone}, C.},
        title = "{Offset multireflector antennas with perfect pattern symmetry and polarization discrimination}",
      journal = {AT\&T Technical Journal},
         year = 1978,
        month = sep,
       volume = {57},
        pages = {2663-2684},
       adsurl = {https://ui.adsabs.harvard.edu/abs/1978ATTTJ..57.2663D}
}

@book{goldsmith,
    author = "{Goldsmith}, P",
    title = "{Quasioptical Systems: Gaussian Beam Quasioptical Propagation and Applications}",
    publisher = "IEEE Press series on microwave technology and techniques" ,
    year = "1998"}

@ARTICLE{spectral_distortions,
       author =  {{Chluba}, J. and {Abitbol}, M.~H. and {Aghanim}, N. and {Ali-Ha{\"\i}moud}, Y. and {Alvarez}, M. and {Basu}, K. and {Bolliet}, B. and {Burigana}, C. and {de Bernardis}, P. and {Delabrouille}, J. and {Dimastrogiovanni}, E. and {Finelli}, F. and {Fixsen}, D. and {Hart}, L. and {Hern{\'a}ndez-Monteagudo}, C. and {Hill}, J.~C. and {Kogut}, A. and {Kohri}, K. and {Lesgourgues}, J. and {Maffei}, B. and {Mather}, J. and {Mukherjee}, S. and {Patil}, S.~P. and {Ravenni}, A. and {Remazeilles}, M. and {Rotti}, A. and {Rubi{\~n}o-Martin}, J.~A. and {Silk}, J. and {Sunyaev}, R.~A. and {Switzer}, E.~R.},
        title = "{New horizons in cosmology with spectral distortions of the cosmic microwave background}",
      journal = {Experimental Astronomy},
         year = 2021,
        month = jun,
       volume = {51},
       number = {3},
        pages = {1515-1554},
          doi = {10.1007/s10686-021-09729-5},
archivePrefix = {arXiv},
       eprint = {1909.01593},
 primaryClass = {astro-ph.CO},
       adsurl = {https://ui.adsabs.harvard.edu/abs/2021ExA....51.1515C}
}

@ARTICLE{CIB,
       author = {{Puget}, J.-L. and {Abergel}, A. and {Bernard}, J.-P. and {Boulanger}, F. and {Burton}, W.~B. and {Desert}, F.-X. and {Hartmann}, D.},
        title = "{Tentative detection of a cosmic far-infrared background with COBE.}",
      journal = {Astronomy \& Astrophysics},
         year = 1996,
        month = apr,
       volume = {308},
        pages = {L5},
       adsurl = {https://ui.adsabs.harvard.edu/abs/1996A&A...308L...5P}
}

@phdthesis{thesexavier,
    author = {{Coulon}, X.},
    title = {Spéctroscopie du ciel : vers la mesure des distorsions spectrales du fond diffus cosmologique},
    school = {Université Paris-Saclay} ,
    year = 2026
}

@INPROCEEDINGS{spiexavier,
       author = {{Coulon}, X. and {Maffei}, B. and {Aghanim}, N. and {Pagano}, L.},
        title = "{Preparing future instrument to high-precision spectroscopy of the cosmic microwave background}",
    booktitle = {Space Telescopes and Instrumentation 2024: Optical, Infrared, and Millimeter Wave},
         year = 2024,
       editor = {{Coyle}, Laura E. and {Matsuura}, Shuji and {Perrin}, Marshall D.},
       series = {Society of Photo-Optical Instrumentation Engineers (SPIE) Conference Series},
       volume = {13092},
        month = aug,
          eid = {1309239},
        pages = {1309239},
          doi = {10.1117/12.3019817},
       adsurl = {https://ui.adsabs.harvard.edu/abs/2024SPIE13092E..39C}
}

@misc{carmen,
  title = {USER MANUAL FOR CNES ZERO PRESSURE BALLOONS},
  author = {{CNES}},
  year = {2018},
  howpublished = {\url{https://cnes.fr/sites/default/files/2024-07/7_cnes_ballons_-_user_manual_zpb_-_2018_en.pdf}},
  note = {Accessed: 2026-03-22}
}

@misc{FOSSIL,
  title = {FTS fOr CMB Spectral diStortIon expLoration: A mission concept for the M-class ESA call},
  author = {Aghanim, N. and {FOSSIL Collaboration}},
  year = {2022},
  howpublished = {\url{http://www.ias.u-psud.fr/sites/default/files/FOSSIL-web_0.pdf}},
  note = {Accessed: 2026-03-22}
}
\bibliographystyle{spiebib} % makes bibtex use spiebib.bst

\end{document}